\documentclass[aps,prl,preprint,groupedaddress]{revtex4-1}

\usepackage{braket}
\usepackage{graphicx}
\usepackage{mathrsfs}
\usepackage{amsmath}

\begin{document}

\title{Quantum mutual information of an entangled state propagating through a fast--light medium}

\author{$^1$Jeremy B. Clark, $^1$Ryan T. Glasser, $^{1,2}$Quentin Glorieux,
$^3$Ulrich Vogl, $^1$Tian Li, $^4$Kevin M. Jones, $^1$Paul D. Lett}

\affiliation{$^1$Quantum Measurement Division, National Institute of Standards and Technology and Joint Quantum Institute, NIST \& the University of Maryland, Gaithersburg, MD 20899\\
$^2$Group of Applied Physics, University of Geneva, Chemin de Pinchat 22, CH-1211 Geneva, Switzerland\\
$^3$Max Planck Institute for the Science of Light,G\"{u}nther--Scharowsky--Str. 1,Building 24, 91058 Erlangen, Germany\\
$^4$Physics Department, Williams College, Williamstown, MA 01267}

\maketitle

\textbf{
Although it is widely accepted that classical information cannot travel faster than the speed of light in vacuum, the behavior of quantum correlations and quantum information propagating through actively--pumped fast--light media has not been studied in detail.
To investigate this behavior, we send one half of an entangled state of light through a gain--assisted fast--light medium and detect the remaining quantum correlations.
We show that the quantum correlations can be advanced by a small fraction of the correlation time while the entanglement is preserved even in the presence of noise added by phase--insensitive gain.
Additionally, although we observe an advance of the peak of the quantum mutual information between the modes, we find that the degradation of the mutual information due to the added noise appears to prevent an advancement of the leading edge.
In contrast, we show that both the leading and trailing edges of the mutual information in a slow--light system can be significantly delayed.
}

Over the past decade, many experiments have demonstrated the ability to manipulate the group velocities of optical pulses moving through atomic vapors \cite{camacho_wide-bandwidth_2007,hau_light_1999, glasser_demonstration_2012,keaveney_maximal_2012}.
In particular, much work has been done to understand fast--light phenomena associated with anomalous dispersion, which gives rise to group velocities that are greater than the speed of light in vacuum (or even negative) \cite{milonni_fast_2010}.
For classical pulses propagating without the presence of noise, it has been well established theoretically \cite{1914Brillouin,1907Sommerfeld} that the initial turn--on point of a pulse (the ``pulse front") propagates through a linear causal medium at the speed of light in vacuum.
It is often argued \cite{chiao_vi:_1997} that this signal carries the entirety of the pulse's classical information content since the remainder of the pulse can in principle be inferred by measuring the pulse height and its derivatives just after the point of non-analyticity has passed.

Experimentally, particularly in the inevitable presence of quantum noise, pulse fronts may not convey the full story of what is readily observed in the laboratory.
It is thus interesting to consider other operational definitions of a signal that apply to particular systems.
For example, Stenner, et al. \cite{stenner_speed_2003} studied the propagation of classical information encoded in bright, actively--shaped optical pulses traveling through a fast--light medium.
These experiments showed that the operational information velocity is actually slowed to speeds less than $c$.
Although noise may have affected the experimental results, these experiments were not conducted in a regime where quantum noise necessarily played a crucial role.
On the other hand, adopting a definition of signal velocity based on observing a given signal--to--noise ratio, Kuzmich, et al. showed how quantum noise associated with gain--assisted fast--light would be expected to limit the early detection of smooth, narrowband pulses consisting of only a few photons \cite{kuzmich_signal_2001}.

\begin{figure}
\centering\includegraphics[width=.9\columnwidth]{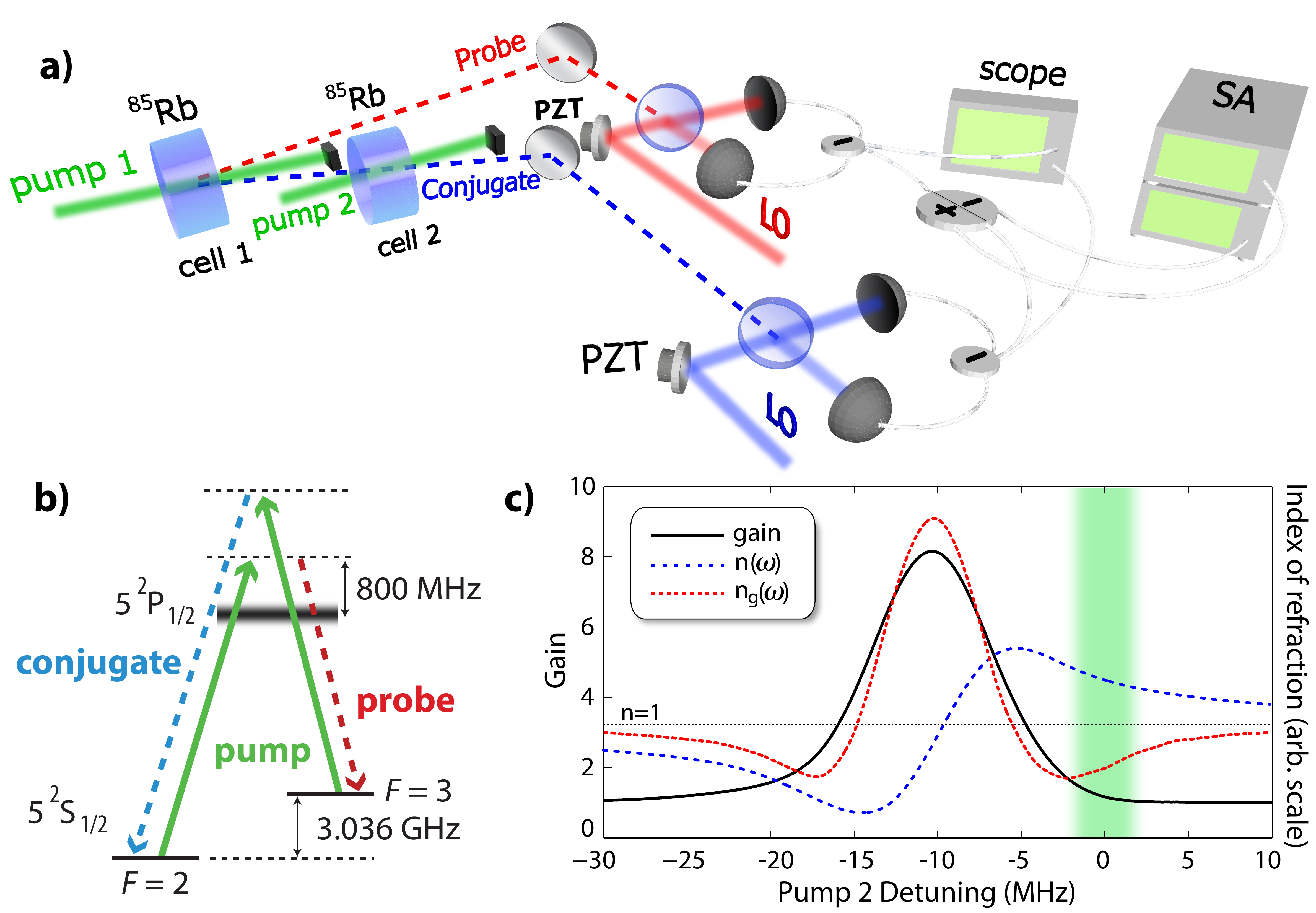}
\caption{\label{fig:setup} a) Experimental set-up.  Vacuum-squeezed twin beams are generated in cell 1 using 4WM in a double-lambda configuration (pictured in part b).
We create a region of anomalous dispersion for the conjugate in a second vapor cell using a second 4WM process driven by pump 2, whose frequency is independently tunable with respect to pump 1 (see supplemental material).
We scan the phase of the local oscillators (LOs) using piezo-electric transducers (PZT) in order to verify the presence of entanglement.
The sum and difference signals of the homodyne detections are recorded on a pair of spectrum analyzers (SAs) to detect quantum correlations.
An oscilloscope is triggered to detect time traces of the individual homodyne detectors given a predetermined threshold of squeezing heralded by the spectrum analyzers.
c) Measured gain profile (black solid line) of the second 4WM process as a function of the detuning of pump 2 relative to pump 1.
From the gain profile, we numerically compute the associated refractive index $n(\omega)$ and group index $n_g$.
In determining the advancement, we confine our attention to fluctuations in the frequency band (shaded green) where we observe quantum correlations generated in cell 1.
We tune the second pump frequency so that the bandwidth of anomalous dispersion coincides with the bandwidth where we observe quantum correlations.
}
\end{figure}

Here we adopt an alternative definition of a signal by choosing it to be the random, but strongly correlated quantum fluctuations between two spatially--separated parts of a bipartite entangled state.
The entangled state in this experiment is generated via four-wave mixing (4WM) in a warm vapor of $^{85}$Rb \cite{boyer_entangled_2008}, which converts two photons from a strong pump beam into ``twin" photons emitted into spatially separated modes referred to as the probe and the conjugate (Fig.~\ref{fig:setup}a).
The fluctuations of the probe and conjugate electric fields are not externally imposed, and they present no obvious pulse fronts or non--analytic features to point to as defining the signal velocity.
As such, classically--rooted approaches to defining the signal or information content of the individual modes are not readily applicable to this system.

Despite the randomness of these fluctuations, however, there is quantum information shared between the modes due to the entanglement.
Although entanglement cannot be used to signal superluminally \cite{Peres_relativity}, it is thought to be an essential resource in quantum information science \cite{Braustein_review, weedbrook_gaussian_2012}.
Accordingly, the prospect of storing \cite{lvovsky_optical_2009} or delaying \cite{marino_tunable_2009} entanglement has attracted significant interest.
To our knowledge, the behavior of entanglement upon propagation through fast--light media has not been characterized experimentally.
Here we investigate the behavior of the mutual information of an entangled state of vacuum-squeezed twin beams using quantum information formalism.
We study how the anomalous dispersion associated with phase--insensitive gain \cite{Boyd2009} affects these correlations by inserting a second vapor cell into the path of the conjugate and driving a second 4WM process with a separate pump (Fig.~\ref{fig:setup}b).
We show that when one portion of the state passes through this fast--light medium, the peak of the quantum mutual information between the modes is advanced, but the arrival of the leading edge is not.
We also show that, in contrast, the leading and trailing edges of the mutual information are both delayed when one of the modes propagates through a slow--light medium with the same level of added noise.

The real and imaginary parts of the nonlinear susceptibility $\chi^{(3)}$ that govern the response of the second 4WM process to the conjugate can be described by a set of equations similar to the Kramers-Kronig relations applicable to linear dielectric media \cite{hutchings_nonlinear}.
These relations stipulate that the gain profile of the medium will generally give rise to a frequency band of anomalous dispersion \cite{Boyd2009} on the wings of the gain line.
Using this relationship between gain and dispersion as a guide (see Fig.~\ref{fig:setup}c), we change the detuning of the pump beam used to drive the fast--light 4WM process such that the conjugate frequency overlaps with the region of anomalous dispersion.
Since the frequency of the conjugate beam is approximately 4 GHz to the blue of the atomic absorption line, blocking the second pump renders the atomic vapor in the second cell nearly dispersionless and non-absorptive.
Thus, comparing transmission of the conjugate through the cell with the pump unblocked or blocked permits a comparison between propagation through a fast--light medium or free space, respectively.   

By performing separate balanced homodyne detections of the probe and conjugate modes, we measure the fluctuations of the in-phase ($\hat{X}$) and out-of-phase ($\hat{Y}$) amplitudes of the electromagnetic field in each beam, which are referred to as the field quadratures.
Taken individually, the probe and conjugate beams exhibit quadrature fluctuations that exceed the shot noise limit.
Taken together, however, these fluctuations display strong correlations beyond the limits achievable classically.
To characterize the strength of the correlations, it is helpful to introduce the joint quadrature operators $\hat{X}_-=(\hat{X}_p-\hat{X}_c)/\sqrt{2}$ and $\hat{Y}_+=(\hat{Y}_p+\hat{Y}_c)/\sqrt{2}$ where $p$ and $c$ denote the probe and conjugate fields, respectively.
For the appropriate choice of local oscillator phases \cite{boyer_entangled_2008}, the fluctuations of one of the joint quadratures ($\braket{\Delta\hat{X}_-^2}$ or $\braket{\Delta\hat{Y}_+^2}$) fall below the shot noise limit (are ``squeezed").

We verify the presence of entanglement by calculating a related quantity, the inseparability ($\mathcal{I}$):
\begin{equation}
\mathcal{I}\equiv\langle \Delta \hat{X}^2_-\rangle_m+\langle \Delta \hat{Y}^2_+\rangle_m.
\end{equation} 
Here $\langle \Delta \hat{X}^2_-\rangle_m$ is the minimum value of the difference signal, $\langle \Delta \hat{Y}^2_+\rangle_m$ is the minimum value of the sum, and each term is normalized to the shot noise limit. 
An inseparability $\mathcal{I}<2$ is a necessary and sufficient condition to conclude that any bipartite Gaussian state is entangled \cite{duan_inseparability_2000}.
We measure the inseparability by allowing the local oscillator phases to drift and  triggering an oscilloscope to record time traces of each individual homodyne detector when the local oscillator phases are chosen to observe the squeezing in one joint quadrature (see supplementary material and Fig.~\ref{fig:setup}).
We then add or subtract the homodyne time traces (according to which joint quadrature is squeezed), Fourier transform the result, and integrate the power spectral density over a 100~kHz--2~MHz bandwidth.
By acquiring noise power statistics for both joint quadratures, we calculate the inseparability.

\begin{figure}
\centering\includegraphics[width=0.7\columnwidth]{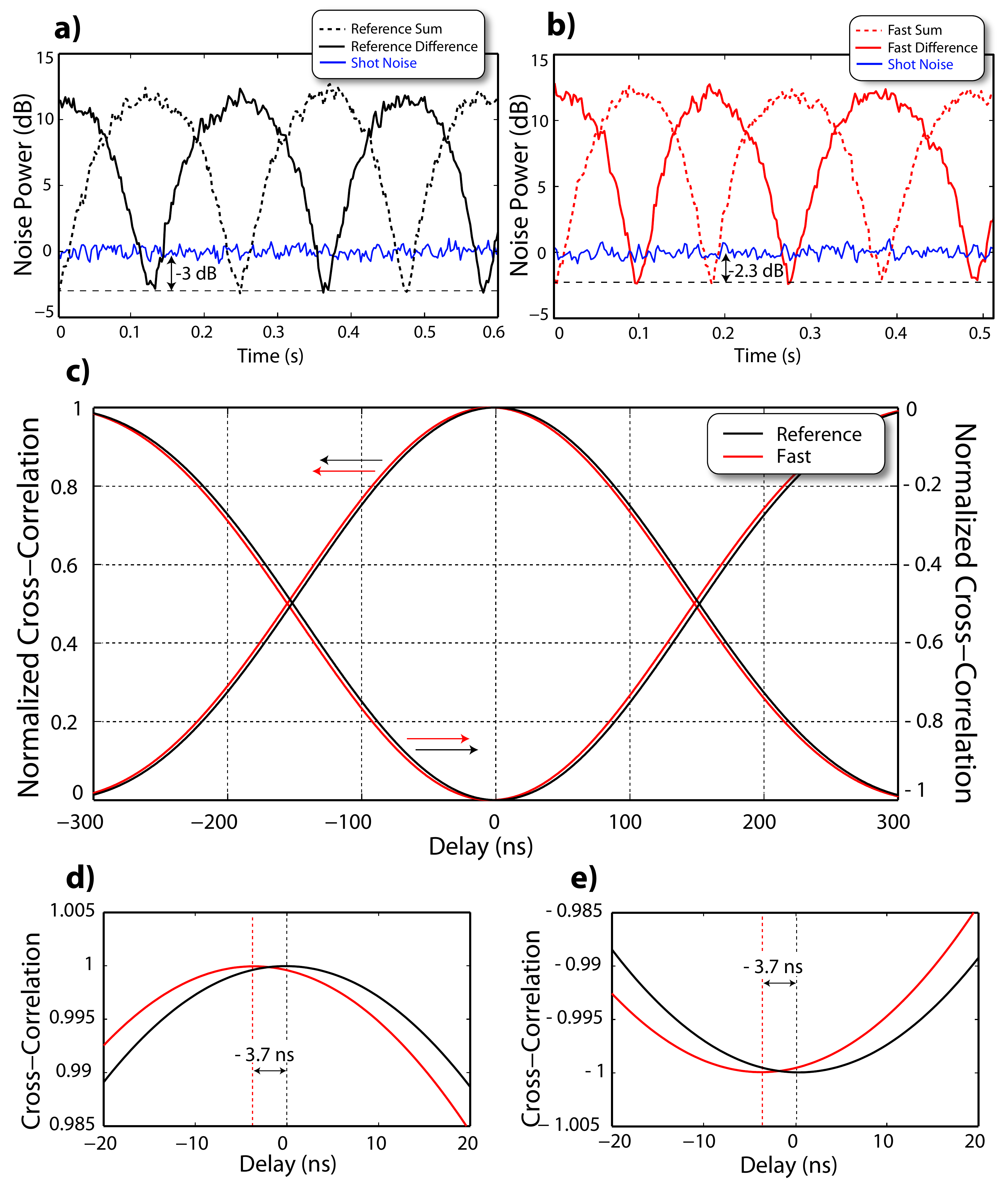}
\caption{\label{fig:sqz_advance}
Persistence of correlations associated with entanglement in the presence of anomalous dispersion.
a) We observe up to -3 dB of squeezing with an associated inseparability $\mathcal{I} \approx 1$ when the second (fast--light) four-wave mixing process is suppressed.
b) In the presence of a small phase--insensitive gain giving rise to anomalous dispersion, the squeezing reduces to -2.3 dB and $\mathcal{I}$ increases to 1.2, which is still sufficient to show entanglement ($\mathcal{I}<2$).
c) Average normalized cross-correlation functions for the correlated and anti-correlated joint quadratures.
The reference and fast--light data are both subject to a low-pass filter (supplementary information) used to match the frequency band of the cross-correlation function to that used to calculate $\mathcal{I}$.
Parts d) and e) provide closer looks at the peak correlation and anti-correlations, respectively.
}
\end{figure}

Studies of bright beam propagation through fast--light media \cite{vogl_advanced_2013} have investigated the trade--off between the magnitude of advancement and the amount of added noise \cite{caves_quantum_1982,boyd_noise_2010} as a function of detuning.
Here we choose a detuning of the second pump that produces a readily detectable advancement of the conjugate fluctuations without significantly deteriorating the inseparability.
By operating in a regime of low gain (G~$\approx$~1.1), we maintained an inseparability of $\mathcal{I}=1.2$ under fast--light conditions, confirming the persistence of entanglement between the probe and conjugate after the conjugate passes through the fast--light medium (Fig.~\ref{fig:sqz_advance}a).

Other experiments involving bright classical pulses propagating through 4WM--based fast--light media showed that lower seed intensities lead to smaller advancements \cite{glasser_stimulated_2012}, making it unclear whether it would be possible to detect the advancement of a few--photon state of light.
A standard practice to demonstrate an advance or delay in the arrival of a pulse peak traveling through a dispersive medium is to scale the maximum of the output pulse to match that of the input (since dispersion is typically accompanied by gain or loss).
Analogously, we confirm that the fluctuations of the continuous--wave conjugate are advanced through the fast--light cell by computing the normalized cross--correlation function of the detected probe and conjugate quadratures for both the reference and fast--light cases (Fig.~\ref{fig:sqz_advance}c--e).
We evaluate the correlation functions after filtering the probe and conjugate homodyne time traces with a 100 kHz--2 MHz bandpass filter (see supplementary information) to insure that the cross-correlation only uses fluctuations in the frequency range that was used to evaluate the inseparability, $\mathcal{I}$.
After acquiring 200 time traces, we conclude that the peak of the cross-correlation function is shifted forward in time by $3.7\pm~0.1~$ns, corresponding to a fractional advance of $\approx1\%$ relative to the cross-correlation width (approximately 300~ns).
Here, the uncertainty is estimated by taking the standard deviation of the mean for the cross--correlation peak advancements over all the experiments.

While useful to clearly see an advancement of the correlations, the normalized cross--correlation function of the field quadratures does not capture how the noise added through phase--insensitive gain of the fast--light 4WM process affects the transport of any quantum information.
As highlighted by Kuzmich, et al. \cite{kuzmich_signal_2001}, understanding the effects of added noise is critical since they could play a vital role in limiting effective signal speeds.
We studied the strength of the entanglement as a function of the relative delay by plotting (Fig.~\ref{fig:advance}a) the average inseparability, $\mathcal{I}$, as a function of the relative delay (see supplementary information).
The delay is implemented in software in exactly the same manner as when calculating the cross--correlation function.
Although the minimum value of $\mathcal{I}$ is advanced in time for the fast--light case, its degradation acts, within the experimental uncertainty, to prevent the leading edge from advancing forward in time.
Figure \ref{fig:advance}b provides a sampling of the delay--dependent squeezing measurements used to calculate the inseparability, which indicates an advance in the maximum squeezing of 3.7~$\pm$~0.1~ns (Fig.~\ref{fig:advance}c).

In our experiment, where we are measuring the continuous random fluctuations of the probe and conjugate beams, there is no imposed ``signal" as such.
The fluctuations on one beam, however, carry information about the fluctuations on the other.
We can capture this by calculating the quantum mutual information between the two beams (Fig.~\ref{fig:mutual_info}), working from the same basic data as used to calculate the delay--dependent inseparability.
The mutual information quantifies the total (classical plus quantum) correlations between the probe and conjugate \cite{ollivier_quantum_2001}.
To calculate the mutual information of our twin beam state, we exploit the fact that any bipartite Gaussian state can be completely characterized by the variances and covariances of the field quadratures \cite{serafini_symplectic_2004}.
By assuming that the state of the twin beams is Gaussian, we are able to measure the mutual information by retrieving relevant statistical moments of the quadrature fluctuations from the photocurrent time traces \cite{vogl_experimental_2013}.
In good agreement with the squeezing and cross-correlation measurements, we observe an advancement of $3.7~\pm~0.1$~ns of the peak of the delay--dependent mutual information, paired with a degradation due to uncorrelated noise added by the fast--light cell (see supplementary information).
This degradation appears to prevent us from observing an advance of the leading edge of the fast--light mutual information (red curve in Fig.~\ref{fig:mutual_info}).

While superluminal information velocities would violate Einstein causality, subluminal information velocities do not.
This observation has led to the misconception that slow group velocities might necessarily limit the propagation speed of classical information in slow--light media.
In a follow--up study involving classical pulses in slow--light media, Stenner, et al. found \cite{stenner_fast_2005} that the velocity of classical information should be viewed as distinct from the group velocity for slow--light pulses.
Moreover, it has been suggested \cite{chiao_vi:_1997, Parker200423} that the set of ``non--analytic" points in physical waveforms should be thought of as the only carriers of classical information, which by bandwidth arguments must travel through the medium precisely at c.
Under this interpretation, dispersive media in general do not affect the propagation of classical information.

Given these classically--rooted interpretations, it might seem plausible to expect that if the fast--light medium in our experiment were replaced with a slow--light medium, the delayed mutual information would lie within the envelope defined by the reference case (no delay).
Using the techniques outlined in \cite{marino_tunable_2009}, we tuned the temperature and the pump detuning of the fast--light 4WM process to slow the propagation of the probe.
We slowed the probe to the greatest extent possible while limiting the degradation of the inseparability to the same level as with the fast--light case.
The behavior of the delay--dependent slow--light mutual information is plotted alongside the reference and fast--light cases (green trace in Fig.~\ref{fig:mutual_info}).
Given an equivalent degradation of the quantum mutual information with added noise, we are able to observe significant delays of the leading and trailing edges of the mutual information compared to the reference case.
Additionally, we observe broadening of the full-width at half-maximum of the delayed mutual information, which is generally consistent with the character of distortion expected in slow--light systems \cite{boyd_slow_2007}.
These results highlight behavioral differences between fast and slow light in the quantum regime.

To summarize, we have demonstrated an advancement in time of the quantum fluctuations in one mode of an entangled state of light passing through a fast--light medium while preserving the entanglement between the modes.
We showed that the peak of the quantum mutual information between the modes can be advanced in time, but added noise associated with the dispersion prevents us from observing an advance of the leading edge.
In contrast, in a slow--light medium operating under conditions which produce a similar reduction in the peak mutual information, the leading and trailing edges of the mutual information can be significantly delayed.
We hope that this work will motivate further inquiry into the propagation of quantum information through fast and slow--light media.

This research was supported by the Physics Frontiers Center at the Joint Quantum Institute and the Air Force Office of Scientific Research. QG performed this work with the support of the Marie Curie IOF FP7 Program - (Multimem - 300632), UV with the support of the Alexander von Humboldt Foundation, and RG with the support of a National Research Council Research Associateship Award at NIST.
\bibliography{fast_entangle_bib}

\newpage

\newpage

\newpage
\begin{figure}
\centering\includegraphics[width=\textwidth]{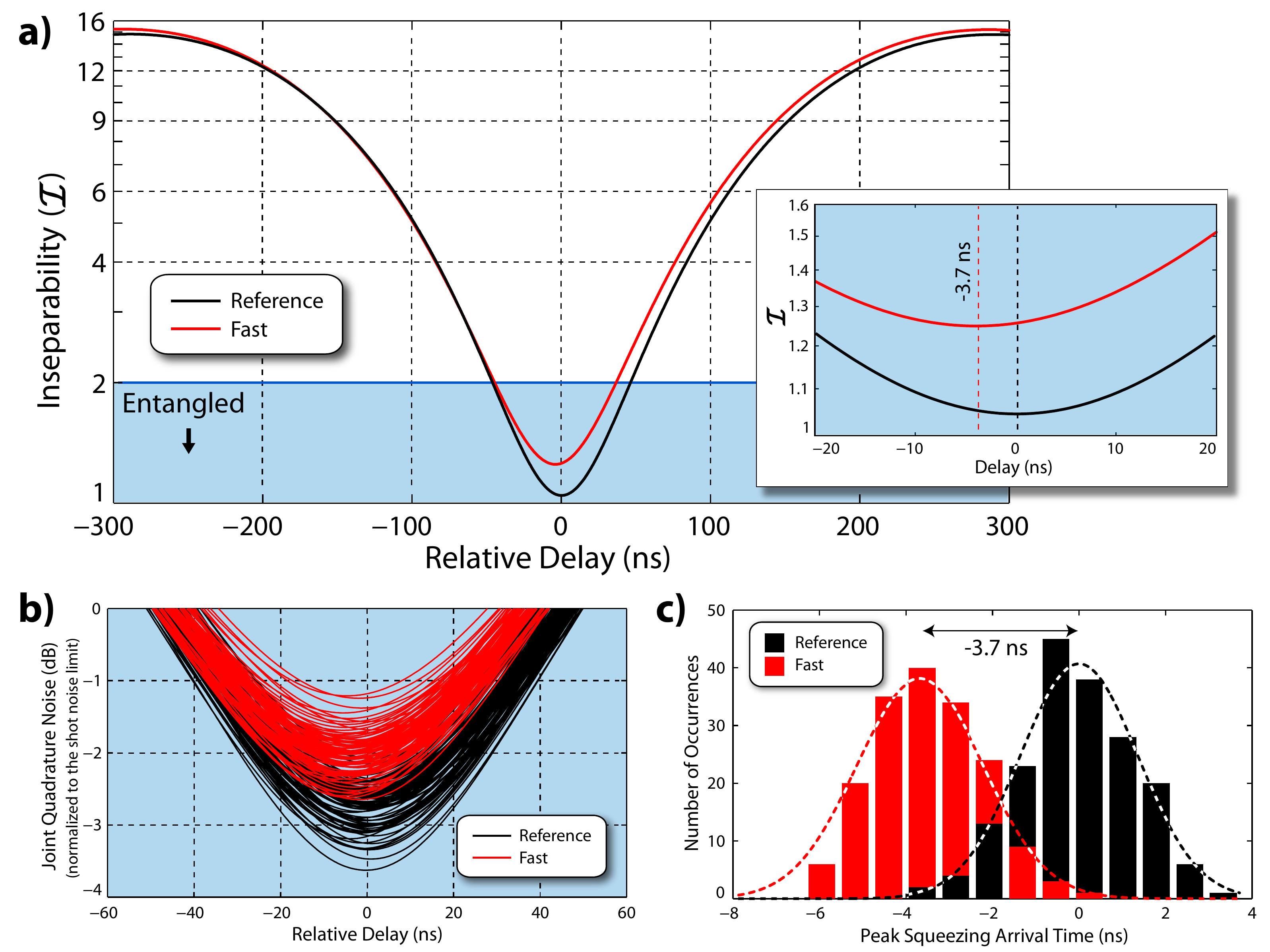}
\caption{\label{fig:advance}
a) Average advance and accompanying degradation of the inseparability, $\mathcal{I}$ ($\mathcal{I}<2$ implies entanglement), in the presence of anomalous dispersion (fast, red curves) and upon blocking the second pump (reference, black curves).
b) Sampling of the squeezing versus delay over 200 experimental iterations used to compute the average $\mathcal{I}$.
c) Histogram of the sampled minima of the joint quadrature noise (i.e. maximum squeezing) versus the relative probe--conjugate delay.
From the sampled shots we extract an advance of 3.7~$\pm$~0.1 ns where the uncertainty has been estimated by computing the standard deviation of the mean.
}
\end{figure}

\newpage
\begin{figure}
\centering\includegraphics[width=\columnwidth]{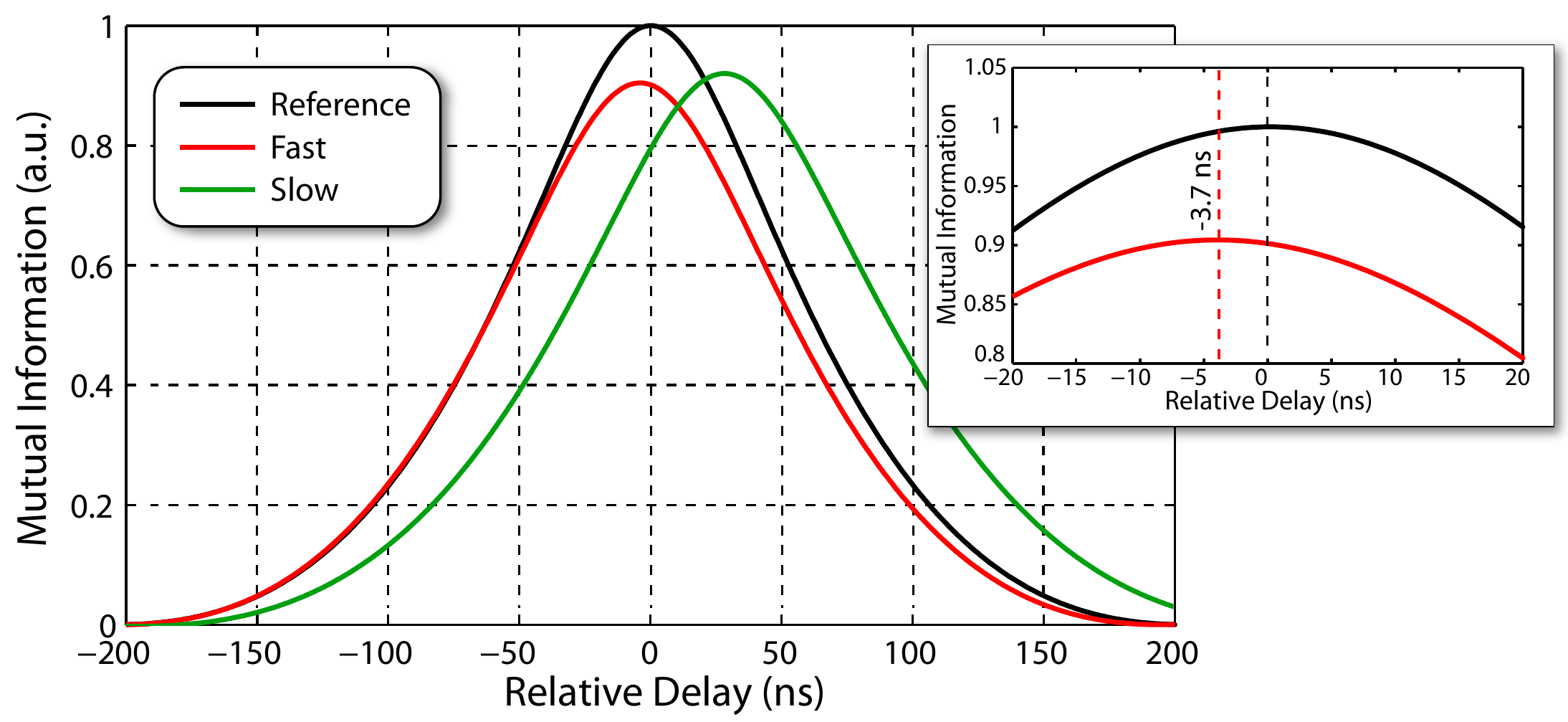}
\caption{\label{fig:mutual_info}
Comparison of the computed quantum mutual information between the probe and conjugate as a function of the relative delay for the cases of fast and slow light.
When considering the fast--light advancement of the conjugate (red trace), we observe an advance in the peak of the mutual information of 3.7~$\pm$~0.1~ns.
There is no statistically significant advance of the leading edge of the mutual information in the case of fast--light propagation.
Repeating the same analysis for slow--light propagation of the probe, we observe a delay of both the leading and trailing edge of the mutual information by hundreds of standard deviations (green trace).
}
\end{figure}

\clearpage

\begin{center}
\textbf{{\Large Supplementary Information}}
\end{center}

\bigskip
Technical details related to the generation and detection of squeezing and entanglement in our system have been explained at length elsewhere [S1].
Here we develop in more detail the techniques used to repeatably achieve the same dispersive properties in the fast--light four--wave mixing process over the course of many measurements, a necessary condition for gathering statistics.
We also provide further details regarding the detection settings and data analysis used to show an advance in the quantum correlations.
We briefly describe the data required to compute the quantum (von Neumann) mutual information, which is based on [S2, S3].
Finally, we conclude by presenting a simple theoretical model showing how the inseparability parameter $\mathcal{I}$ and von Neumann mutual information $I(P;C)$ are affected by the introduction of a phase--insensitive gain to one of the twin beams.

\section{Laser System}
In order to repeatably manipulate the dispersive properties of the $^{85}$Rb vapor in the fast--light vapor cell, it was essential that the small relative detuning (tens of MHz) between the two pumps be stable and easily tunable to within $\approx$~100~kHz.
To achieve these requirements, we used two double-passed acousto-optic modulators (AOMs) to shift the frequency of approximately 10~mW of light picked off from a master Ti:sapphire laser, which itself was stabilized to a reference cavity (Fig.~S1).
When oppositely driven, the cascaded double-passed AOMs enabled us to tune the second pump beam relative to the first by up to $\pm$100~MHz.
Double--passing the AOMs was necessary in order to decouple the shift in frequency from the direction of propagation so that the shifted light could be used to injection--lock a slave diode.

The output of the slave diode laser was amplified using a tapered amplifier and spatially filtered using single--mode fiber (SMF).
Seeding the slave diode with the light shifted by the AOMs ensured sufficient optical power ($\approx$~30~mW) to saturate the tapered amplifier and maintain a stable pump power over the tuning range.
After amplification and spatial filtering, we were left with 300~mW of optical power.
The mode was gently focused to a waist of approximately 1.5~mm inside of the second cell (which was heated to 105~$^\circ$C) used to create the region of anomalous dispersion.
Once the lock was established, we could tune the master Ti:sapphire laser several GHz without the slave diode losing the lock or observing any change in optical power of the fast--light pump.

\renewcommand{\figurename}{Fig.~S1}
\renewcommand{\thefigure}{}
\begin{figure}
\centering\includegraphics[width=.8\textwidth]{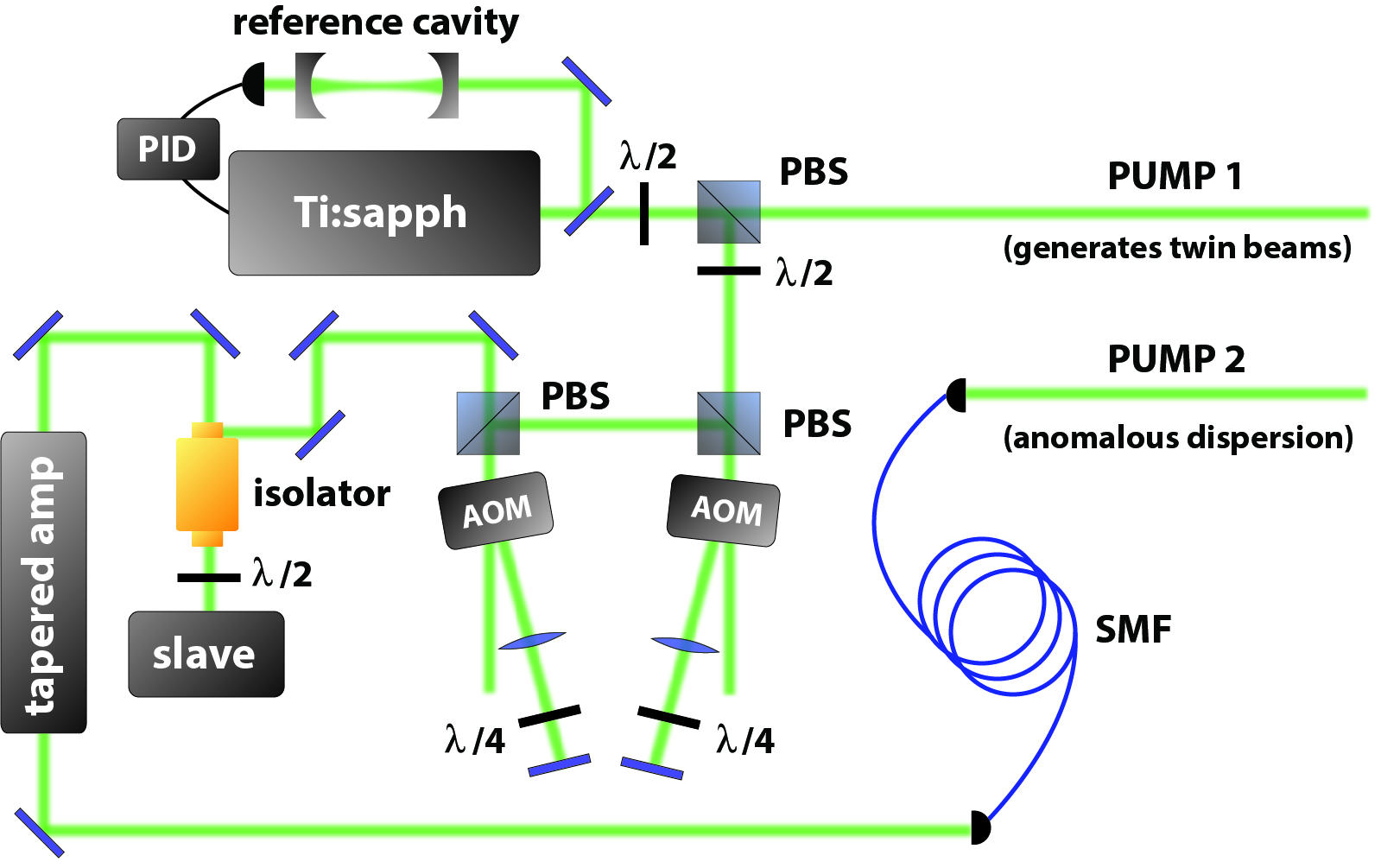}
\caption{\label{fig:secondpump} Preparation of the fast--light pump.  10~mW of power is picked off from a Ti:sapphire laser (frequency--stabilized to a reference cavity) using a half--wave plate ($\lambda/2$) and a polarizing beamsplitter (PBS).  This light was then coupled into two acousto-optic modulators (AOMs) arranged in oppositely--shifted double--pass configurations.  The output of the second AOM was used to injection lock a slave diode laser, and the slave output was amplified using a tapered amplifier.  The spatial mode of this shifted, amplified light was spatially filtered using a single-mode fiber (SMF) before being used to drive the second (fast--light) four-wave mixing process.}
\end{figure}

\section{Data Acquisition}
In order to acquire time traces of the two homodyne detections, we allowed the phases of the local oscillators to drift while we synchronously triggered two spectrum analyzers and an oscilloscope.
The spectrum analyzers detected the noise power present on the sum and difference signals of the homodyne detections while the oscilloscope was configured to measure the direct time--dependent output of each homodyne detector.
The spectrum analyzers were configured to measure a central frequency of 750~kHz with a resolution bandwidth of 30~kHz and a sweep time of 10~ms.
Although the oscilloscope was also triggered at the same time as the spectrum analyzers, the time traces were only saved upon measuring an average of at least -2~dB of squeezing on either spectrum analyzer.
The traces were sorted by whether they corresponded to the local oscillator phases giving rise to squeezing in $\hat{X}_-$ or $\hat{Y}_+$.
The oscilloscope traces consisted of 1 million points acquired at a sampling frequency of 2.5~gigasamples per second for a total acquisition time of 400~$\mu$s.

We acquired 100 time traces of the homodyne detections of the twin fields for the reference and fast--light cases (yielding a total of 200 traces).
The data acquisition time typically lasted a total of about 45 minutes.
In between these recordings, we took 5 measurements of shot noise by blocking the vacuum-squeezed twin beams and integrating the power spectral density over the 100~kHz--2~MHz frequency range.
No temperature or detuning settings were changed during any given set of measurements.
We repeated this procedure over the course of several days to verify reproducibility.

\section{Analysis}
The first step in the analysis was to determine which detection (optical sideband) frequencies exhibited quantum correlations indicative of entanglement ($\mathcal{I}<2$).
While scanning the local oscillators, we detected the squeezing of $\braket{\Delta \hat{X}_-^2}$ or $\braket{\Delta\hat{Y}_+^2}$ at different detection frequencies (with a constant resolution bandwidth of 30~kHz throughout).
Given any detected squeezing, we calculated the inseparability by summing the normalized noise powers of $\braket{\Delta \hat{X}_-^2}$ and $\braket{\Delta\hat{Y}_+^2}$ on a linear scale to calculate $\mathcal{I}$ (see Fig.~S2).
We observed the strongest squeezing (and therefore smallest values of $\mathcal{I}$) at a central detection frequency of 750~kHz.
More generally, we observed $\mathcal{I}<2$ at detection frequencies ranging from 20~kHz to 3~MHz.
Accordingly, we selected a detection frequency of 750~kHz to trigger data acquisitions with the oscilloscope.

\begin{figure}
\renewcommand{\figurename}{Fig.~S2}
\renewcommand{\thefigure}{}
\centering\includegraphics[width=.9\textwidth]{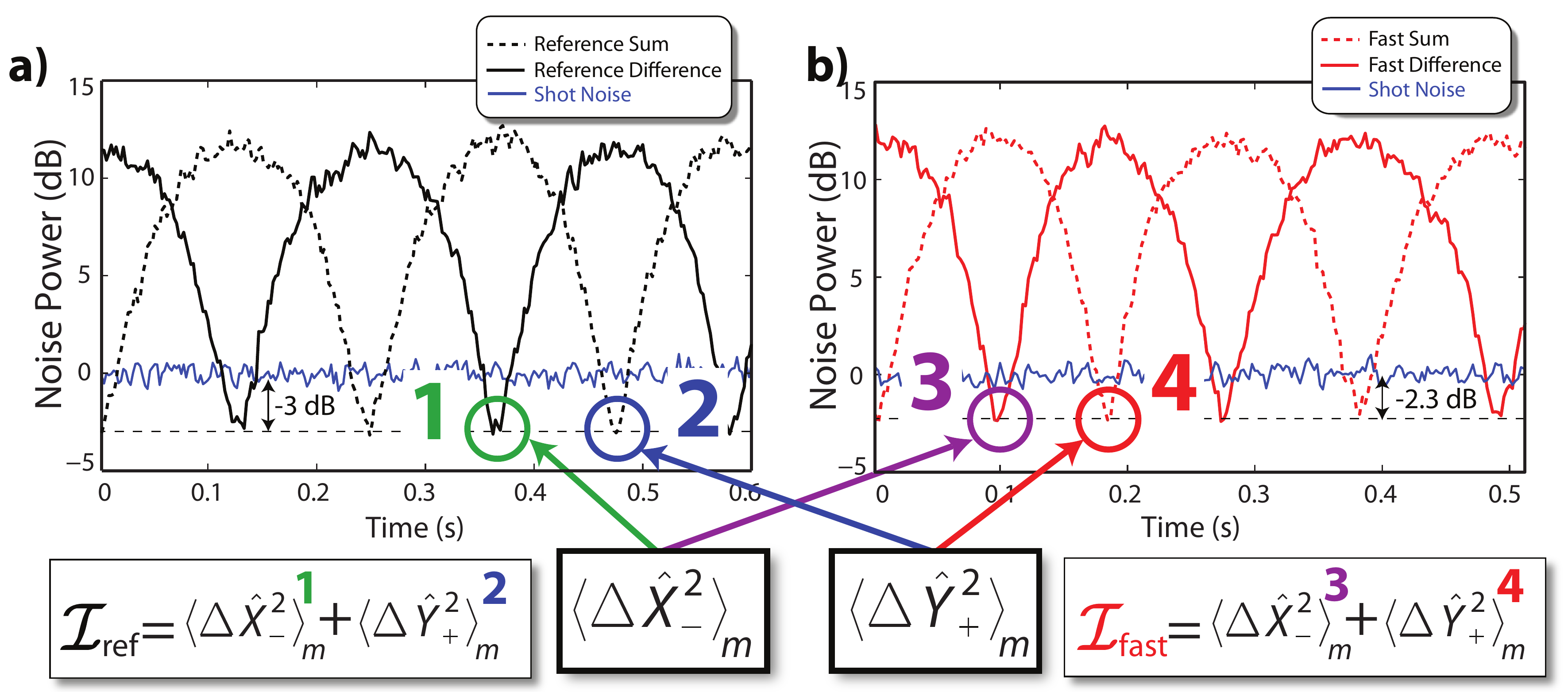}
\caption{\label{fig:inseparability} Determination of inseparability $\mathcal{I}$ using the spectrum analyzer (center frequency 750~kHz, resolution bandwidth 30~kHz, sweep time 1~s). 
Part a) corresponds to the reference case while b) corresponds to data taken with fast light.
The minimum noise power of the homodyne difference signal corresponds to squeezing of the $\hat{X}$ quadrature difference (i.e. $\braket{\Delta \hat{X}_-^2}<1$) while the minimum noise power of the homodyne sum gives the squeezing of phase sum $\braket{\Delta\hat{Y}_+^2}$.
We observed the strongest squeezing (lowest value of $\mathcal{I}$) when detecting at 750~kHz.
With the fast--light cell inserted into the path of the conjugate, we were able to see $\mathcal{I}<2$ (assuming a resolution bandwidth of 30~kHz throughout) at central detection frequencies ranging from 20~kHz to approximately 3~MHz.
The figure shows how the inseparability can be calculated (from the data obtained from the spectrum analyzers) by adding together the values of the appropriate minima and averaging.
}
\end{figure}

\subsection{Measuring the Cross--Correlation}
For two real discrete functions $f[n]$ and $g[m]$ (as is the case in this experiment), the discrete cross--correlation function $(f\star g)[n]$ is defined according to
\begin{equation}
(f\star g)[n] \equiv \sum\limits_{m=-\infty}^\infty f[m]g[n+m].
\end{equation}
For real experiments, the limits on the sum are bounded from above by the length of the data sets (in this case, our time traces consist of 10 million points).
Since we only observed entanglement at detection frequencies from $\approx$~20~kHz up to $\approx$~3~MHz, we were careful to filter out any frequencies where we could not definitively show quantum correlations when computing the temporal cross-correlation function between the beams.
In other words, we wanted to prevent any frequencies exhibiting two-mode excess noise from accounting for any possible advance in the cross--correlation function (since we are interested in studying the propagation of quantum information associated with entanglement).
Figure~S3 illustrates the band-pass window that we applied to the probe and conjugate homodyne data prior to computing the cross--correlation function.

\begin{figure}
\renewcommand{\figurename}{Fig.~S3}
\renewcommand{\thefigure}{}
\centering\includegraphics[width=.8\textwidth]{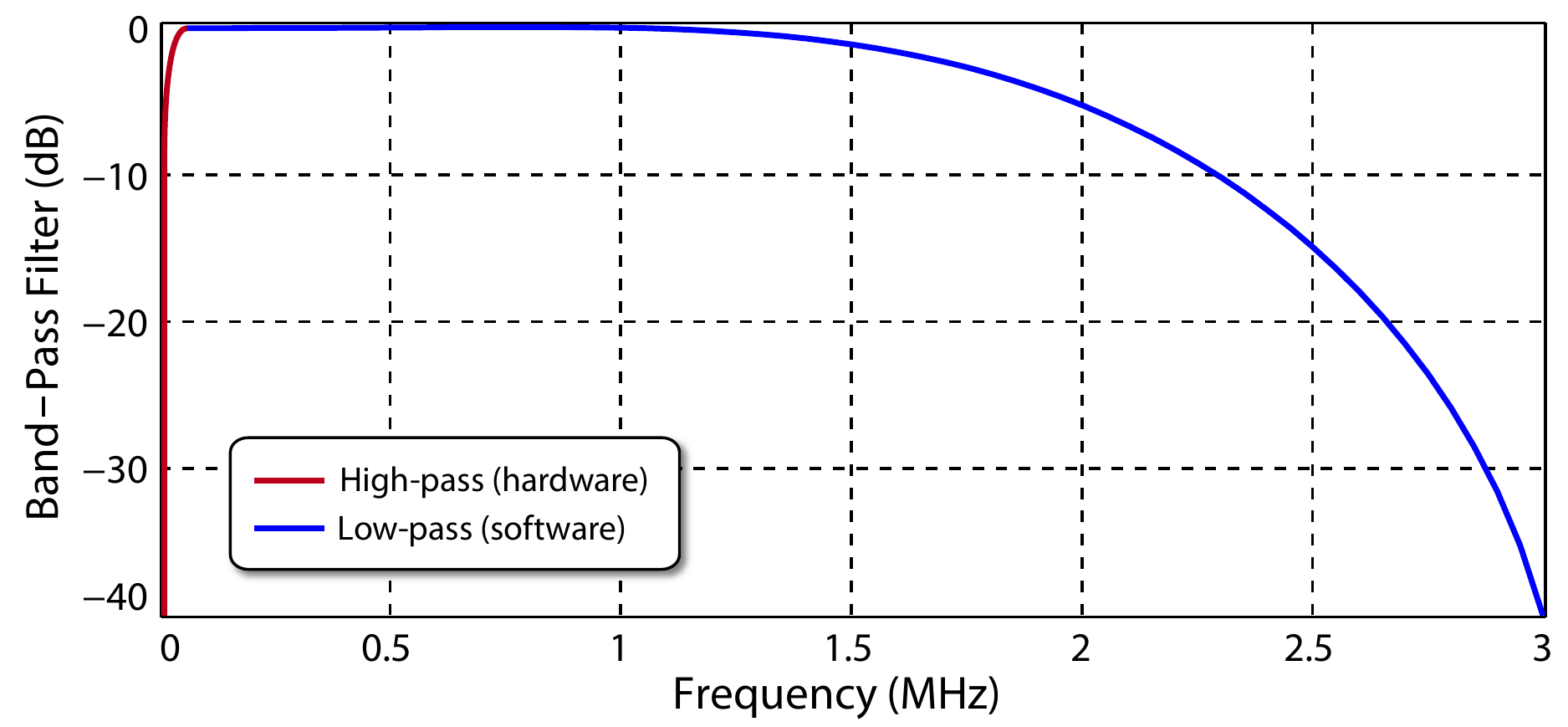}
\caption{\label{fig:hanning} The frequency profile of the effective band-pass filter that was used to filter the probe and conjugate homodyne detections for all three cases (fast--light medium, slow--light medium, and the reference case).  
The high-pass portion of the filter (at frequencies below 100~kHz, red curve) was achieved in hardware using an LC circuit before the homodyne time traces were recorded on the oscilloscope.
The attenuation of the higher frequencies was achieved in software using a Hanning window with a -3~dB cutoff point of 1.75~MHz.}
\end{figure}

\subsection{Measuring the Inseparability}
To study the advance in the squeezing and inseparability, we computed the Fourier transform of the sum or difference of the probe and conjugate homodyne detections (depending on whether the oscilloscope was triggered by detecting squeezing of $\braket{\Delta \hat{X}_-^2}$ or $\braket{\Delta \hat{Y}_+^2}$).
We then integrated the resulting power spectrum from 100~kHz--2~MHz, well within the bandwidth where we observed entanglement.
Selecting a wide bandwidth ensured a large total noise power after the integration, which assisted in repeatably obtaining the same magnitude of advancement.
We repeated this process when detecting shot noise in order to normalize the joint noise power obtained with the probe and conjugate.
We then shifted the probe and conjugate time traces in time (in steps of 0.4~ns, the sampling period of the oscilloscope) before repeating this procedure over a range of relative delays.

\subsection{Measuring the Mutual Information}
To compute the quantum mutual information, we assume that the initial two--mode squeezed state is Gaussian (i.e. it is completely described by a Gaussian Wigner function).
Furthermore, we assume that the state remains Gaussian after one portion propagates through the second 4WM process (the fast/slow--light cell).
In other words, we assume that only Gaussian quantum operations take place in either of these cells. 
These assumptions simplify the calculations significantly since any Gaussian state is fully characterized by the first and second moments of the field quadratures, which can be obtained from measurements taken with homodyne detectors.

To obtain the excess noises of the individual fields, we Fourier transformed the time traces of the probe and conjugate homodyne detections and integrated the power spectral density over the same 100~kHz--2~MHz bandwidth used to measure the squeezing and inseparability.
The two--mode squeezing and the excess noises of the individual beams and are sufficient to compute the state's covariance matrix, which is given by
\begin{equation}
\gamma_{ij} = \frac{1}{2}\langle\hat{R}_{i}\hat{R}_{j}+\hat{R}_{j}\hat{R}_{i}\rangle-\langle\hat{R}_{i}\rangle\langle\hat{R}_{j}\rangle.
\end{equation}
where $R_{i}\equiv(\hat{X}_{i},\hat{Y}_{i})$ and $i\in\{p,c\}$ ($p$ denotes the probe and $c$ the conjugate).
In this standard form, the on-diagonal sub-matrices characterize the individual modes' fluctuations while the off-diagonal sub-matrices capture the covariances between the two modes' quadrature fluctuations [S3].
In a similar fashion to the delay--dependent squeezing, we computed the delay--dependent covariance matrix of our state.

The quantum mutual information, also referred to as von Neumann mutual information, is defined under quantum information theory [S4].
For a general two--mode state this quantity is defined in terms of the von Neumann entropy $S_V(\rho)=-\mathrm{Tr}(\rho \log\rho)$ according to
\begin{equation}
I(\rho)=S_V(\rho_1)+S_V(\rho_2)-S_V(\rho).
\end{equation}
Here $\rho$ denotes the full state density matrix and $\rho_{i}$ denotes the reduced density matrix of the subsystems after the partial trace has been evaluated over the other mode.
For the case of a continuous--variable Gaussian state, the calculation of the mutual information involves the symplectic eigenvalues of the standard--form and partially-transposed covariance matrix, as described in more detail elsewhere [S2, S3].
From the delay--dependent covariance matrix of the state of vacuum--squeezed twin beams, we computed the average delay--dependent quantum mutual information.

The uncertainty in determining the advance of the peak of the mutual information for the fast--light case ($\pm~0.1$~ns) was determined by computing the standard deviation of the peak advancements over all 200 experiments.
To estimate the uncertainty of the leading edge, we computed the standard deviation of the mean of the magnitude of the mutual information at a given delay time and then used standard error propagation techniques to estimate the uncertainty in the timing of the leading edge.
This led to an estimated time uncertainty of no more than 0.7~ns in the leading edge for relative delays that are more positive than (to the right of) -150~ns.
For relative delays ranging from -200~ns to -150~ns, the uncertainty approaches 1.5~ns.

\section{Quantum Mutual Information under Fast-- and Slow--light Conditions}
While superluminal information velocities would violate Einstein causality, subluminal information velocities do not.
This observation has sometimes led to the misconception that slow group velocities might necessarily limit the propagation speed of classical information in slow-light media.
In a follow-up study to [S6] involving classical pulses in slow-light media, Stenner, \textit{et al.} found [S7] that the velocity of classical information should be viewed as distinct from the group velocity for slow-light pulses.
Moreover, it has been suggested [S8, S9] that the set of ``non-analytic" points in physical waveforms should be thought of as the only carriers of classical information, which by bandwidth arguments must travel through the medium precisely at $c$.
Under this interpretation, dispersive media in general do not affect the propagation of classical information.

It might seem plausible to expect that if the fast-light medium in our experiment were replaced with a slow-light medium, the delayed mutual information would again lie within the envelope defined by the reference case.
Using the techniques outlined in [S10], we tuned the temperature and the pump detuning of the fast-light 4WM process to slow the propagation of the probe (note that we use the probe beam in this case, instead of the conjugate, as we do for fast light.
The mutual information, as defined in Eq. (3), should be the same for the two cases).
We slowed the probe to the greatest extent possible while limiting the degradation of the inseparability to the same level as with the fast-light case.
The gain in the slow light cell was approximately 1.2 and the detuning of the probe from the peak of the 4WM gain in the second cell was less than 1 MHz (near line center, neglecting the light shift).
The 1-photon detuning for the 4WM process was approximately 800 MHz, as in the fast-light case.
The behavior of the delay-dependent slow-light mutual information is plotted alongside the reference and fast-light cases in Fig. 4 of the manuscript (green trace).
Given an equivalent degradation of the quantum mutual information with added noise, we are able to observe significant delays of the leading and trailing edges of the mutual information compared to the reference case.
This is strikingly different from the apparent restriction on advancing the mutual information in the fast-light case.
In addition, while the traditional method of looking at non-analytic points in classical signals results in an invariant transport of information at the vacuum speed of light, $c$, looking at the quantum mutual information yields different results for fast and slow light.

\section{Effects of Phase--Insensitive Gain}
\begin{figure}
\renewcommand{\figurename}{Fig.~S4}
\renewcommand{\thefigure}{}
\centering\includegraphics[width=\textwidth]{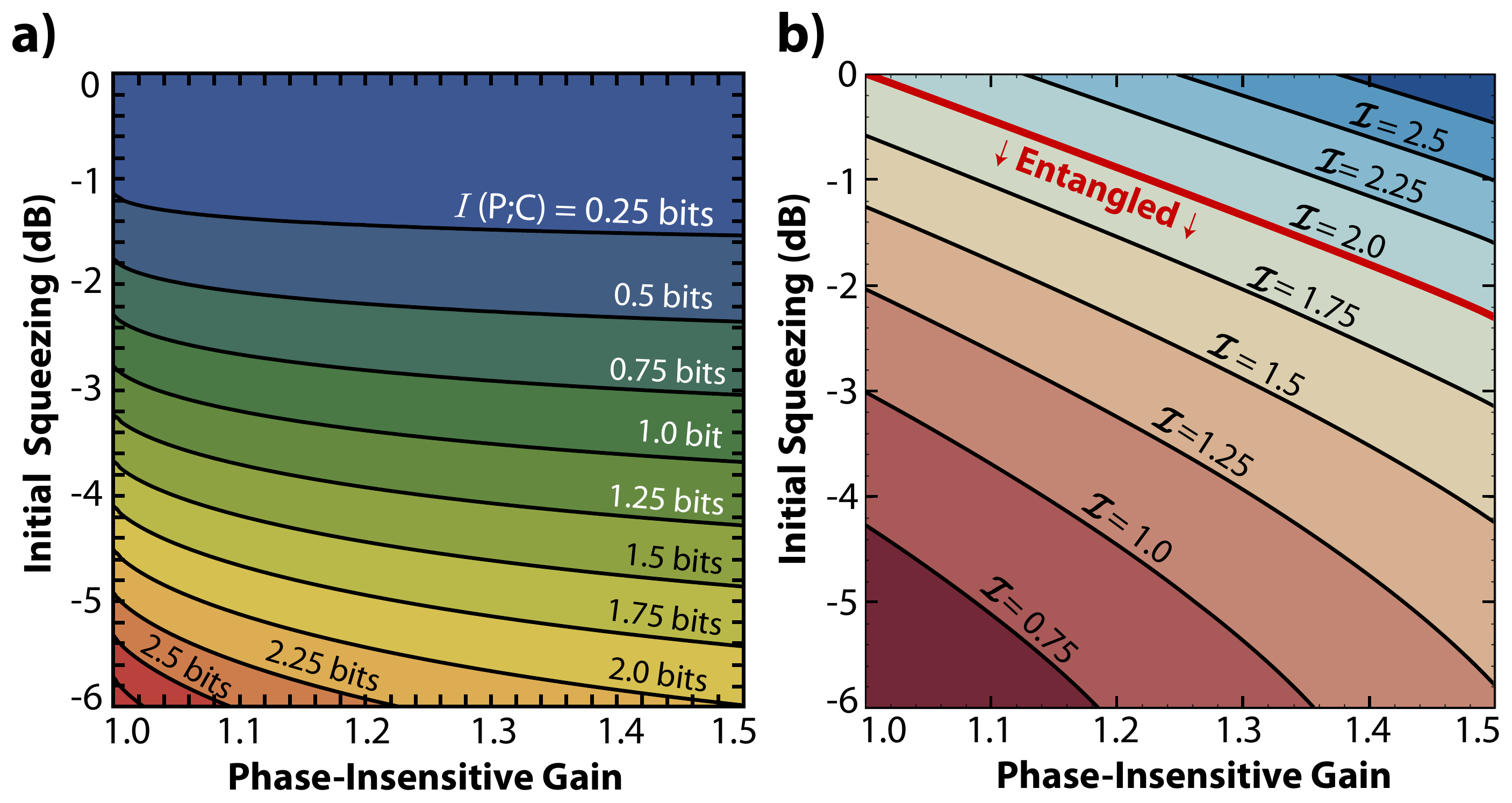}
\caption{\label{fig:mutual_info_decay} a) Decay of the von Neumann mutual information $I(P;C)$ and b) growth of the inseparability parameter $\mathcal{I}$ when one of the twin beams is subject to a phase-insensitive gain.
In this model, the probe and conjugate are assumed to initially be in a pure EPR state.
The units in part a) reflect the assumption that the squeezed noise power in the detection bandwidth $\mathcal{B}$ is flat and sampled for a time $\Delta t=\mathcal{B}^{-1}$.
The red demarcation in part b) indicates the total phase--insensitive gain required to lose entanglement given some initial two--mode squeezing ($\mathcal{I}<2~\longleftrightarrow$~entanglement).
}
\end{figure}
We provide a rudimentary analysis of how the phase--insensitive gain associated with the fast--light (slow--light) medium would be expected to affect the mutual information, $I(P;C)$, and inseparability parameter, $\mathcal{I}$.
A more complete treatment might incorporate the dispersive properties of the fast--light (slow--light) medium and the squeezed power spectrum to obtain any delay--dependent behavior of the system.
Our objective here is only to establish a simple theoretical picture of how the peak mutual information and minimum inseparability would be expected to behave (given certain initial conditions) after one of the entangled beams is sent through a phase--insensitive amplifier.

Assuming that the initial two--mode squeezed vacuum state is ideal (i.e. a pure EPR state), the state's covariance matrix assumes the form
\begin{center}
$\boldsymbol{\gamma}=\left( \begin{array}{cccc}
\cosh2r & 0 & \sinh2r & 0\\
0 & \cosh2r & 0 & -\sinh2r\\
\sinh2r & 0 & \cosh2r & 0\\
0 & -\sinh2r & 0 & \cosh2r\\
\end{array}\right)$
\end{center}
where $r$ is the squeezing parameter and the level of squeezing is given by $e^{-2r}$.
It is well known that passing one mode of this state through a phase--insensitive amplifier will necessarily add noise.
The effect of phase-insensitive gain can be described by the transformation [S5]
\begin{equation}
\label{eq:transform}
\hat{a}\rightarrow\mu \hat{a}+\nu \hat{b}^\dagger.
\end{equation}
Here $\hat{a}$ denotes the annihilation operator of one of the twin modes and $\hat{b}^\dagger$ the creation operator for the second, unused (i.e. vacuum--seeded) input port of the amplifier.
The coefficients are related to the phase-insensitive gain $G$ according to $|\mu|^2=G$ and $|\nu|^2=G-1$.
When one of the two EPR modes is subject to an ideal phase--insensitive amplifier, the covariance matrix of the resulting state can be calculated to be
\begin{center}
$\boldsymbol{\gamma}=\left( \begin{array}{cccc}
\cosh2r & 0 & \sqrt{G}\sinh2r & 0\\
0 & \cosh2r & 0 & -\sqrt{G}\sinh2r\\
\sqrt{G}\sinh2r & 0 & 2G\cosh^2r-1 & 0\\
0 & -\sqrt{G}\sinh2r & 0 & 2G\cosh^2r-1\\
\end{array}\right).$
\end{center}
The von Neumann mutual information of the twin beams can be calculated from this covariance matrix as a function of the gain $G$ and the initial two--mode squeezing. 
As illustrated in Fig.~S4a, phase--insensitive amplification of one of the modes leads to a monotonic decay of the von Neumann mutual information $I(P;C)$ as a function of the gain.
A similar procedure can be used to evaluate the amplifier's effect on the inseparability parameter $\mathcal{I}$, which goes as

\begin{equation}
\mathcal{I}=(1+G)\cosh2r-2\sqrt{G}\sinh2r+(G-1).
\end{equation}
Figure~S4b shows how the introduction of gain leads to an increase in the inseparability parameter $\mathcal{I}$ and, eventually, to a loss of bipartite entanglement ($\mathcal{I}<2$ is a necessary and sufficient condition to conclude that any Gaussian state is entangled).
\begin{center}
\line(1,0){250}
\end{center}
\noindent
[S1] Vincent Boyer, Alberto M. Marino, Raphael C. Pooser, and Paul D. Lett.  Entangled images from four-wave mixing. \textit{Science} \textbf{321}, 544--547 (2008).
\\
\\
\noindent
[S2] Alessio Serafini, Fabrizio Illuminati, and Silvio De Siena. Symplectic invariants, entropic measures and correlations of Gaussian states. \textit{J.~Phys.~B.} \textbf{37}, L21--L28 (2004).
\\
\\
\noindent
[S3] Ulrich Vogl, \textit{et al.}. Experimental characterization of Gaussian quantum discord generated by four-wave mixing. \textit{Phys.~Rev.~A} \textbf{87}, 010101 (2013).
\\
\\
\noindent
[S4] Mark M. Wilde. From classical to quantum Shannon theory. arXiv:1106.1445 (2011).
\\
\\
\noindent
[S5] Carlton Caves.  Quantum limits on noise in linear amplifiers. \textit{Phys. Rev. D} \textbf{26}, 1817--1839 (1982).
\\
\\
\noindent
[S6] M. D. Stenner, D. J. Gauthier, and M. A. Neifield.  \textit{Nature} \textbf{425}, 695--698 (2003).
\\
\\
\noindent
[S7] M. D. Stenner, D. J. Gauthier, and M. A. Neifeld.  \textit{Phys. Rev. Lett.} \textbf{94}, 053902 (2005).
\\
\\
\noindent
[S8] R. Y. Chiao and A. M. Steinberg, in \textit{Progress in Optics} \textbf{37}, edited by E. Wolf (Elsevier, 1997) pp. 345--405.
\\
\\
\noindent
[S9] R. W. Boyd and P. Narum.  \textit{J. Mod. Opt.} \textbf{54}, 2403--2411 (2007).
\\
\\
\noindent
[S10] A. M. Marino, R. C. Pooser, V. Boyer, and P. D. Lett.  \textit{Nature} \textbf{457}, 859--862 (2009).

\end{document}